\documentclass[
reprint,
amsmath,
amssymb,
]{revtex4-1}
\usepackage{graphicx}

\begin{document}

\title{Gain spectroscopy of a type-II VECSEL chip}

\author{C.~\surname{Lammers}}
\email{christian.lammers@physik.uni-marburg.de}
\author{M.~\surname{Stein}}
\author{C.~\surname{Berger}}
\author{C.~\surname{M\"{o}ller}}
\author{C.~\surname{Fuchs}}
\affiliation{Department of Physics and Material Sciences Center, Philipps-Universit\"{a}t Marburg, Renthof 5, 35032 Marburg, Germany}
\author{A.~\surname{Ruiz Perez}}
\affiliation{$\mathrm{NAsP}_{\mathrm{III/V}}$ GmbH, Hans-Meerwein-Stra\ss{}e, 35032 Marburg, Germany}
\author{A.~\surname{Rahimi--Iman}}
\affiliation{Department of Physics and Material Sciences Center, Philipps-Universit\"{a}t Marburg, Renthof 5, 35032 Marburg, Germany}
\author{J.~\surname{Hader}}
\author{J.~V.~\surname{Moloney}}
\affiliation{College of Optical Sciences, University of Arizona, 1630 E. University Blvd., Tucson, AZ 85721, USA}
\author{W.~\surname{Stolz}}
\author{S.~W.~\surname{Koch}}
\author{M.~\surname{Koch}}
\affiliation{Department of Physics and Material Sciences Center, Philipps-Universit\"{a}t Marburg, Renthof 5, 35032 Marburg, Germany}

\date{\today}

\begin{abstract}
Using optical pump--white light probe spectroscopy the gain dynamics is investigated for a VECSEL chip which is based on a type-II heterostructure. 
The active region the chip consists of a GaAs/(GaIn)As/Ga(AsSb)/(GaIn)As/GaAs multiple quantum well. 
For this structure, a fully microscopic theory predicts a modal room temperature gain at a wavelength of 1170 nm, which is confirmed by experimental spectra. 
The results show a gain buildup on the type-II chip which is delayed relative to that of a type-I chip. 
This slower gain dynamics is attributed to a diminished cooling rate arising from reduced electron--hole scattering. 
\end{abstract}


\maketitle
Very recently, laser operation has been demonstrated in a type-II vertical-external-cavity surface-emitting laser (VECSEL).\cite{Moeller} The type-II design promises further wavelength flexibility in the infrared as well as reduced Auger losses.\cite{Bewley:2008jq,Meyer:1998bz}
This makes lasers with a type-II based gain medium promising for many applications, such as optical data transmission.\cite{murphy2010semiconductor}

Important aspects of the optical properties and carrier dynamics of type-II quantum film structures have been investigated already 25 years ago.\cite{Meissner:1991jm,Binder:1991ks,Galbraith:1992gs,Feldmann:1989im}    
Lately, type-II semiconductor heterostructures with a ``W''-type band alignment of the conduction band, spatially separating  electron and hole confinement, were found promising for laser applications as reported in Refs.~\onlinecite{Berger:2015ev,Gies:2015ec}.

Up to date, both, edge emitters as well as surface-emitting quantum well lasers based on type-II material systems have been demonstrated.\cite{Klem,:/content/journals/10.1049/ip-opt_19982304} 
The gain properties of such structures are often characterized by  stationary methods such as the Hakki--Paoli technique or the variable-stripe-length method.\cite{Hakki:1973fm,Shterengas:2013dl,Shaklee:1973hi,Kirstaedter:1996fe}
Information on the gain dynamics associated with the carrier dynamics inside the laser medium can be obtained via ultrafast pump--probe experiments.\cite{Hall:1990dv,Giessen:1996db,Lange:2007ek}

Here, we use optical pump--white light probe spectroscopy to study the gain dynamics of a type-II ``W''-VECSEL chip containing 10 x (GaIn)As/Ga(AsSb) ``W''-multiple quantum wells.
A fully microscopic theory predicts significant modal room temperature gain for this chip at a wavelength around 1170\,nm.
For comparison, we additionally study a conventional type-I chip.
We find that the gain build up is delayed in the type-II structure as compared to the type-I chip.
We attribute this slower dynamics to a reduced carrier cooling rate in the type-II structure.

\begin{figure}[!t]
  \includegraphics[width=8.5cm]{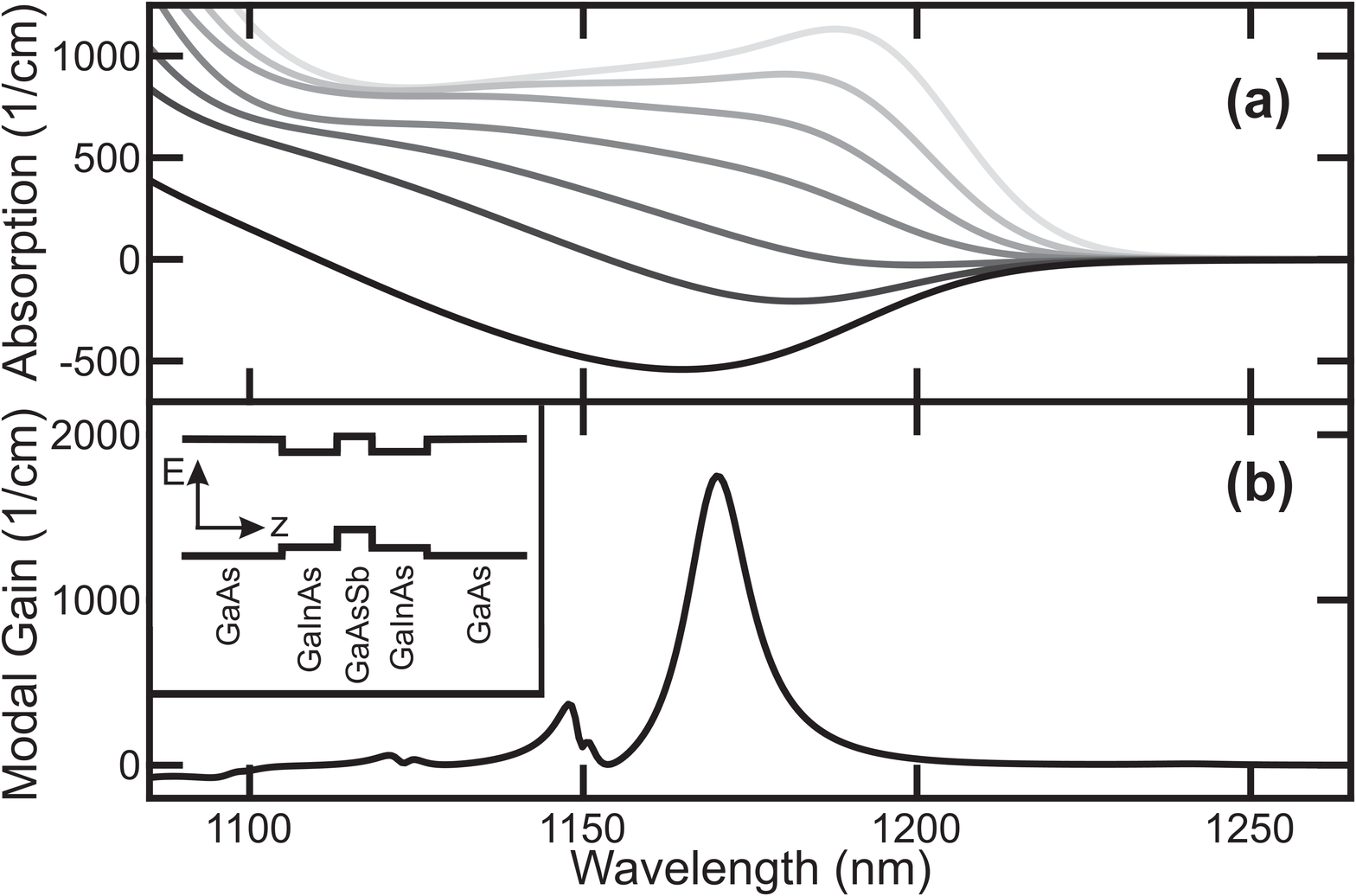}
  \caption{
    Theoretical material absorption and modal gain.
    (a) Calculated material absorption for the active region of the type-II ``W''-laser structure is presented for carrier densities ranging from $0.1\cdot10^{12}/\text{cm}^2$ up to $3\cdot10^{12}/\text{cm}^2$ (bright to dark).
    (b) A simulation of the modal gain for the full type-II VECSEL system is plotted for a carrier density of $3\cdot10^{12}/\text{cm}^2$. The inset illustrates schematically the band alignment of the type-II heterostructure.
  }
  \label{fig:fig1}
\end{figure}

Our VECSEL chip was grown bottom-up on exact GaAs (001) ($\pm$\,0.1$^{\circ}$) substrate in a horizontal AIXTRON AIX 200 Gas Foil Rotation (GFR) metal organic vapor phase epitaxy (MOVPE) reactor system.
It consists of a lattice matched (GaIn)P capping layer, a resonant periodic gain (RPG) structure with 10 multi-quantum well heterostructures (MQWHs) and a distributed Bragg reflector (DBR) consisting of 22 1/2 pairs of (AlGa)As/AlAs.
The MQWH consists of a ``W''-shaped (GaIn)As/Ga(AsSb)/(GaIn)As type-II band alignment, which forms the active region, as can be seen in the inset of Fig.~\ref{fig:fig1}~(b).
These active regions are separated by GaAs/Ga(AsP)/GaAs barriers.
In order to obtain a resonant structure, the optical layer thicknesses of the barriers as well as the (GaIn)P capping layer were matched to $\lambda$/2 with respect to the lasing wavelength.
The layer thicknesses and compositions of the active region were determined by fitting a fully dynamical simulation to the HR-XRD ((004)-reflection) pattern of the chip structure.
This analysis yields a $Ga(As_{1-x}Sb_{x})$ (x=0.198) layer thickness of 4.0\,nm.
The $(Ga_{1-y}In_{y})As$ (y=0.203) layer has a thickness of 5.5\,nm.

In the following, we present calculations on the basis of the semiconductor Bloch equations (SBE) to predict the gain properties of this structure under optical excitation.
In particular, we calculate the changes in absorption and refractive index for various carrier densities.\cite{Lindberg:1988hf,Haug:2009tb,Kira:2011up}
To achieve a sufficient level of accuracy in our calculations, we include all terms in the SBE up to the second Born level to take homogeneous broadening into account, i.e. intrinsic microscopic carrier scattering inside the semiconductor.\cite{Chow:1999ts,Hader:2003hi}
Unavoidable and common growth inhomogeneities in such MQWH samples are modeled by an inhomogeneous broadening, a convolution of our theoretical spectra with a Gaussian distribution describing the variation of the bandgap energies.\cite{Hader:2003hm}
Starting point for our calculations are the band structure and the corresponding wavefunctions.
All these single-particle properties are accessed evaluating an $8\times8$ multi band $\mathbf{k}\cdot\mathbf{p}$ model.\cite{Hader:1997gr,Chow:1999ts}
We assume all carriers to be in thermal equilibrium and therefore Fermi distributed in their respective bands.
Due to the type-II design, local charge inhomogeneities can arise.
Consequently, we solve the Schr\"{o}dinger--Poisson equation to obtain the changes to the confinement potential.\cite{Ahn:1988ks}
Based on the band structure and the single-particle wavefunctions, we compute the dipole and Coulomb matrix elements.\cite{Hader:2003hi}

Having modeled the absorption and refractive index changes induced by a given carrier density, we can derive the overall optical properties of the VECSEL structure.
Reflection properties of the sample are calculated using the transfer-matrix method.\cite{Born:1999un,Kira:2011up}
From the transfer-matrix calculations, we obtain a spectral filter function that describes the spatial overlap of the quantum well positions inside the RPG with the intensity maxima of the longitudinal light modes of the optical cavity.\cite{Schafer:2008in}

\begin{figure}[!htb]
  \includegraphics[width=8.5cm]{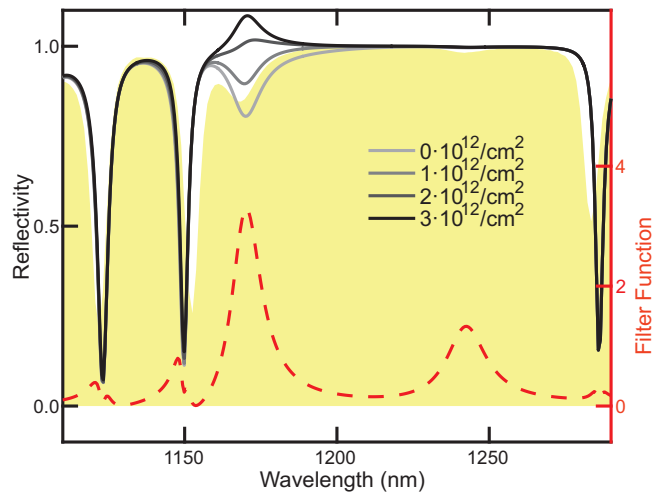}
  \caption{
   Theoretical reflectivity (lines) of the type-II gain structure together with its calculated filter function (dashed line).
   Measured reflectance at room temperature for an unexcited chip is shown as yellow area in the background, while theoretical reflectivity spectra are presented for carrier densities between zero and $3\cdot10^{12}/\text{cm}^2$ (solid lines, bright to dark).
   Again, the system temperature is set to 300\,K with an inhomogeneous broadening of 20\,meV.
  }
  \label{fig:fig2}
\end{figure}

The numerical results for the material absorption of the ``W''-type MQWH at 300\,K are presented in Fig.~\ref{fig:fig1}(a).
Here, we assumed an inhomogeneous broadening of 20\,meV.
As the carrier density is increased from $0.1\cdot10^{12}/\text{cm}^2$ up to $3\cdot10^{12}/\text{cm}^2$ (bright to dark) the excitonic absorption peak vanishes and the absorption becomes negative.
The calculations show that a spectrally broad gain region is formed for this material composition in the RPG.
We compute the modal gain for a carrier density of $3\cdot10^{12}/\text{cm}^2$, by multiplying the filter function (represented as dashed line  in Fig.~\ref{fig:fig2}) with the gain of the RPG structure.
Thereby, the broad material gain shown in Fig.~\ref{fig:fig1}(a) evolves into a narrow modal gain around 1170\,nm.
This is shown in  Fig.~\ref{fig:fig1}(b).

Next, we compare the simulated reflectivity to an experimental spectrum in Fig.~\ref{fig:fig2}.
Here, the measured spectrum plotted in the background of the diagram represents the system under unexcited conditions (yellow shaded area).
A direct comparison between experiment and theory shows a good qualitative agreement.
Using our theory as a predictive tool, we increase the carrier density up to $3\cdot10^{12}/\text{cm}^2$, which leads to the broad material gain, observed in Fig.~\ref{fig:fig1}(a).
Thereby, we observe a gradual increase of the reflectivity at the absorption dip in the stop band.
Finally, the absorption dip evolves into a gain peak around 1170\,nm.

For our optical pump--white light probe gain measurements we use a  1\,kHz, 5\,mJ, 35\,fs regenerative Ti:sapphire amplifier to acquire time-resolved differential reflectivity spectra.
The amplifier drives an optical parametric amplifier (OPA) to provide spectrally tunable fs pump pulses and furthermore generates a fs supercontinuum which is used to probe the gain dynamics in the chip.
The pump spot diameter is determined to 150\,$\mu$m with a knife-edge measurement.
An InGaAs-CCD cooled with liquid nitrogen is used as a detector in combination with a Czerny--Turner monochromator with a 80\,l/mm diffraction grating as dispersive element.
The excitation density of our pump pulse with a central wavelength of 1000\,nm is set to  $5.6\cdot10^{16}/\text{cm}^2$ in order to inject a sufficiently large amount of charge carriers.
All experiments are carried out at room temperature.
To obtain the gain spectra, the measured differential reflectivity spectra were added to the reflectivity measurements of the unexcited sample.
\begin{figure}[!htb]
  \includegraphics[width=8.5cm]{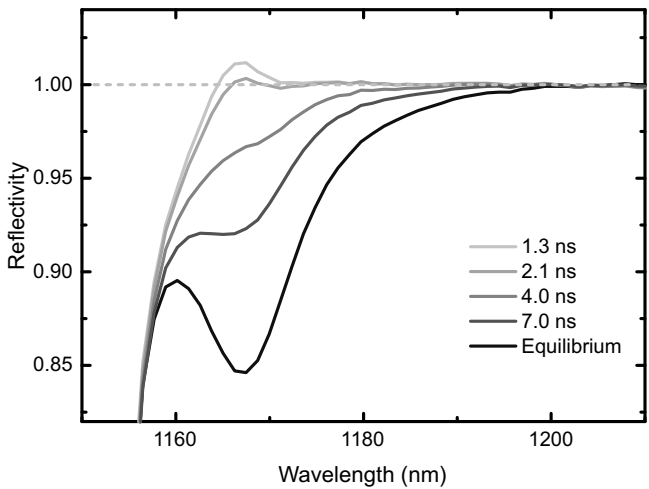}
  \caption{
    Experimental reflectivity spectra of a type-II chip for pump--probe time delays between 1.3 and 7.0\,ns (solid lines, bright to dark).
    The maximum gain is reached 1.3\,ns after excitation.
    The equilibrium situation corresponds to the reflectivity measured by the probe pulse without excitation.
  }
  \label{fig:fig3}
\end{figure}

Examples of the reflectivity spectra obtained for different times after the excitation pulse are shown in Fig.~\ref{fig:fig3}.
For the unexcited sample, one clearly observes the dip in reflection at 1168\,nm which was predicted by theory.
It is governed by an interplay of material absorption with the filter function for the employed material and structure, as Fig.~\ref{fig:fig1} indicates.
After excitation and carrier relaxation the sample shows gain, i.e. the reflectivity exceeds one.
The gain reaches its maximum after 1.3\,ns.
It peaks at 1168\,nm and has a spectral width of 6\,nm.
The gain lasts for several hundred picoseconds.
At 2.1 ns it is barely noticeable.
Thereafter, i.e. for larger delay times, the reflectivity in this spectral region is below one and gradually reduces.
\begin{figure}[!b]
  \includegraphics[width=8.5cm]{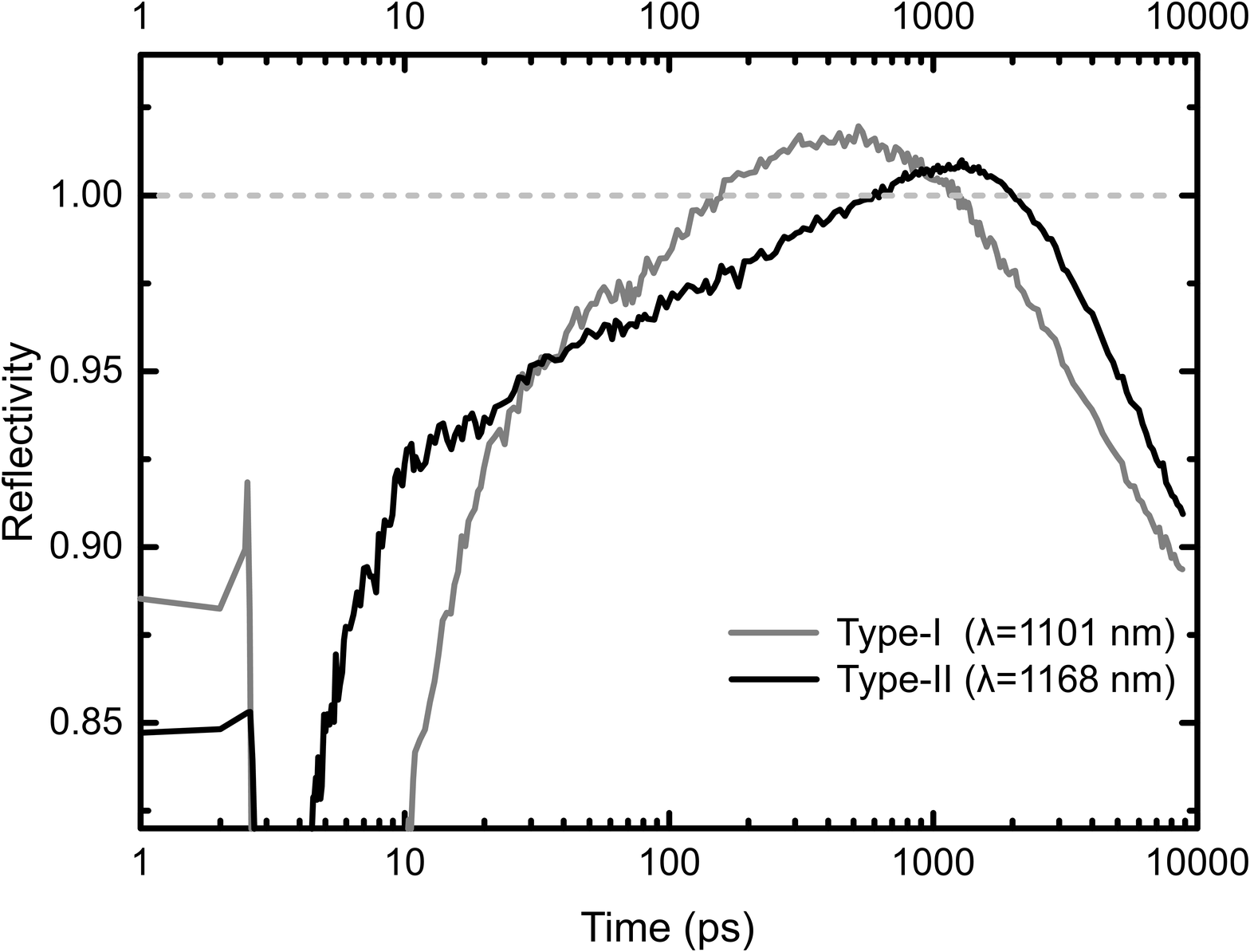}
  \caption{Semilogarithmic plot showing the gain dynamics of the type-II ``W"-VECSEL chip (black line) in comparison to a conventional type-I chip (grey line). For this plot, an offset of 2\,ps was applied to the time axis in order to enable a logarithmic illustration.
  }
  \label{fig:fig4}
\end{figure}

The black curve in Fig.~\ref{fig:fig4} shows the measured gain dynamics of the type-II chip.
To obtain this curve, a region of 1.2 nm width around 1168\,nm was spectrally integrated.
The gain evolves after 600\,ps, reaches its maximum after 1.3 ns and vanishes at 2.1\,ns after the optical excitation.
For comparison, the same measurements were performed on a type-I VECSEL, exploiting spatially direct transitions.
The respective chip, which emits at 1101\,nm, has an RPG comprising 10 (Ga$_{0.71}$In$_{0.29}$)As QWs with a thickness of 5\,nm in between GaAs barriers.
Again, the filter function is positioned at the edge of the stop band.
Hence, all in all one has a similar arrangement as for the type-II VECSEL chip.
The grey curve in Fig.4 shows the gain dynamics of this chip.
Once more, we spectrally integrate over a window of 1.2\,nm around the gain maximum.
In this case, the gain appears at 150 ps and lasts for 1.1\,ns.
By fitting exponential functions to the decreasing reflectivity, we retrieve charge-carrier lifetimes for both VECSEL structures.
These lifetimes amount to 3.3\,ns and 7.3\,ns for the type-I and type-II chip, respectively.
Although the \mbox{type-I} structure produces a higher peak value of 1.016, the time span of gain provided by the material system lasts longer for the type-II structure.
As mentioned above, this can be seen as a direct effect of the spatial separation of charge carriers in type-II MQWHs, which exhibit a reduced recombination rate.
Hence, high charge-carrier densities, which are a prerequisite for gain, are conserved for a longer period of time in comparison to a type-I heterostructure.
It is reasonable to assume that a lower radiative recombination rate in a type-II structure can be a limiting factor with regard to the maximum amount of gain.
We attribute the slower gain build up in the \mbox{type-II} chip to a reduced carrier equilibration rate arising from the following situation: after optical excitation, depending on their excess energies electron and hole distributions emit optical phonons and cool down.
The cooling rates for electrons and holes are typically not identical and differ from material to material.
Yet, in a type-I sample electrons and holes can efficiently exchange energy via Coulomb scattering and hence, the electron and hole distributions have a route to rapidly equilibrate among each other.
If the electron distribution cools slower it can profit from a hole distribution which cools faster - and vice versa.
In a type-II sample electron-hole scattering is significantly diminished due to the reduced spatial overlap of electrons and holes.
Hence, it takes more time until a thermal equilibrium among the two distributions is achieved.
Yet, a microscopic calculation of this effect for the multi-layer structure studied is beyond the scope of this paper.

In conclusion, we have investigated the gain dynamics of a type-II VECSEL chip using optical pump--white light probe spectroscopy.
Experimental gain spectra agree well with a fully microscopic theory.
We observe a slower buildup of the modal gain for the type-II chip when compared to a type-I structure.
We attribute this to a smaller carrier cooling rate in the type-II chip arising from a reduced electron-hole scattering rate.
The smaller scattering rate result from the reduced spatial overlap of electrons and holes wave functions in the type-II design.
Owing to the fact that a type-II chip can be successfully employed in VECSELs and that the design allows for more wavelength flexibility compared to type-I structures, further efforts will be done in upcoming investigations in order to optimize the MQWH design with respect to gain and laser output.
Thereby, type-II VECSELs may push available emission wavelengths further to the highly desired telecom wavelengths.\\


The authors from Marburg acknowledge financial support from the Deutsche Forschungsgemeinschaft via the Collaborative Research Center 1083 (DFG:SFB1083). The authors from Arizona are supported by the AFOSR grant FA9550-14-1-0062.

\bibliography{references}

\begin{thebibliography}{29}%
\makeatletter
\providecommand \@ifxundefined [1]{%
 \@ifx{#1\undefined}
}%
\providecommand \@ifnum [1]{%
 \ifnum #1\expandafter \@firstoftwo
 \else \expandafter \@secondoftwo
 \fi
}%
\providecommand \@ifx [1]{%
 \ifx #1\expandafter \@firstoftwo
 \else \expandafter \@secondoftwo
 \fi
}%
\providecommand \natexlab [1]{#1}%
\providecommand \enquote  [1]{``#1''}%
\providecommand \bibnamefont  [1]{#1}%
\providecommand \bibfnamefont [1]{#1}%
\providecommand \citenamefont [1]{#1}%
\providecommand \href@noop [0]{\@secondoftwo}%
\providecommand \href [0]{\begingroup \@sanitize@url \@href}%
\providecommand \@href[1]{\@@startlink{#1}\@@href}%
\providecommand \@@href[1]{\endgroup#1\@@endlink}%
\providecommand \@sanitize@url [0]{\catcode `\\12\catcode `\$12\catcode
  `\&12\catcode `\#12\catcode `\^12\catcode `\_12\catcode `\%12\relax}%
\providecommand \@@startlink[1]{}%
\providecommand \@@endlink[0]{}%
\providecommand \url  [0]{\begingroup\@sanitize@url \@url }%
\providecommand \@url [1]{\endgroup\@href {#1}{\urlprefix }}%
\providecommand \urlprefix  [0]{URL }%
\providecommand \Eprint [0]{\href }%
\providecommand \doibase [0]{http://dx.doi.org/}%
\providecommand \selectlanguage [0]{\@gobble}%
\providecommand \bibinfo  [0]{\@secondoftwo}%
\providecommand \bibfield  [0]{\@secondoftwo}%
\providecommand \translation [1]{[#1]}%
\providecommand \BibitemOpen [0]{}%
\providecommand \bibitemStop [0]{}%
\providecommand \bibitemNoStop [0]{.\EOS\space}%
\providecommand \EOS [0]{\spacefactor3000\relax}%
\providecommand \BibitemShut  [1]{\csname bibitem#1\endcsname}%
\let\auto@bib@innerbib\@empty
\bibitem [{\citenamefont {M{\"o}ller}\ \emph {et~al.}(2016)\citenamefont
  {M{\"o}ller}, \citenamefont {Fuchs}, \citenamefont {Berger}, \citenamefont
  {Ruiz~Perez}, \citenamefont {Koch}, \citenamefont {Hader}, \citenamefont
  {Moloney}, \citenamefont {Koch},\ and\ \citenamefont {Stolz}}]{Moeller}%
  \BibitemOpen
  \bibfield  {author} {\bibinfo {author} {\bibfnamefont {C.}~\bibnamefont
  {M{\"o}ller}}, \bibinfo {author} {\bibfnamefont {C.}~\bibnamefont {Fuchs}},
  \bibinfo {author} {\bibfnamefont {C.}~\bibnamefont {Berger}}, \bibinfo
  {author} {\bibfnamefont {A.}~\bibnamefont {Ruiz~Perez}}, \bibinfo {author}
  {\bibfnamefont {M.}~\bibnamefont {Koch}}, \bibinfo {author} {\bibfnamefont
  {J.}~\bibnamefont {Hader}}, \bibinfo {author} {\bibfnamefont {J.~V.}\
  \bibnamefont {Moloney}}, \bibinfo {author} {\bibfnamefont {S.~W.}\
  \bibnamefont {Koch}}, \ and\ \bibinfo {author} {\bibfnamefont
  {W.}~\bibnamefont {Stolz}},\ }\href@noop {} {\bibfield  {journal} {\bibinfo
  {journal} {Appl. Phys. Lett.}\ }\textbf {\bibinfo {volume} {108}},\ \bibinfo
  {eid} {071102} (\bibinfo {year} {2016})}\BibitemShut {NoStop}%
\bibitem [{\citenamefont {Bewley}\ \emph {et~al.}(2008)\citenamefont {Bewley},
  \citenamefont {Lindle}, \citenamefont {Kim}, \citenamefont {Kim},
  \citenamefont {Canedy}, \citenamefont {Vurgaftman},\ and\ \citenamefont
  {Meyer}}]{Bewley:2008jq}%
  \BibitemOpen
  \bibfield  {author} {\bibinfo {author} {\bibfnamefont {W.~W.}\ \bibnamefont
  {Bewley}}, \bibinfo {author} {\bibfnamefont {J.~R.}\ \bibnamefont {Lindle}},
  \bibinfo {author} {\bibfnamefont {C.~S.}\ \bibnamefont {Kim}}, \bibinfo
  {author} {\bibfnamefont {M.}~\bibnamefont {Kim}}, \bibinfo {author}
  {\bibfnamefont {C.~L.}\ \bibnamefont {Canedy}}, \bibinfo {author}
  {\bibfnamefont {I.}~\bibnamefont {Vurgaftman}}, \ and\ \bibinfo {author}
  {\bibfnamefont {J.~R.}\ \bibnamefont {Meyer}},\ }\href@noop {} {\bibfield
  {journal} {\bibinfo  {journal} {Appl. Phys. Lett.}\ }\textbf {\bibinfo
  {volume} {93}},\ \bibinfo {pages} {041118} (\bibinfo {year}
  {2008})}\BibitemShut {NoStop}%
\bibitem [{\citenamefont {Meyer}\ \emph
  {et~al.}(1998{\natexlab{a}})\citenamefont {Meyer}, \citenamefont {Felix},
  \citenamefont {Bewley}, \citenamefont {Vurgaftman}, \citenamefont {Aifer},
  \citenamefont {Olafsen}, \citenamefont {Lindle}, \citenamefont {Hoffman},
  \citenamefont {Yang}, \citenamefont {Bennett}, \citenamefont {Shanabrook},
  \citenamefont {Lee}, \citenamefont {Lin}, \citenamefont {Pei},\ and\
  \citenamefont {Miles}}]{Meyer:1998bz}%
  \BibitemOpen
  \bibfield  {author} {\bibinfo {author} {\bibfnamefont {J.~R.}\ \bibnamefont
  {Meyer}}, \bibinfo {author} {\bibfnamefont {C.~L.}\ \bibnamefont {Felix}},
  \bibinfo {author} {\bibfnamefont {W.~W.}\ \bibnamefont {Bewley}}, \bibinfo
  {author} {\bibfnamefont {I.}~\bibnamefont {Vurgaftman}}, \bibinfo {author}
  {\bibfnamefont {E.~H.}\ \bibnamefont {Aifer}}, \bibinfo {author}
  {\bibfnamefont {L.~J.}\ \bibnamefont {Olafsen}}, \bibinfo {author}
  {\bibfnamefont {J.~R.}\ \bibnamefont {Lindle}}, \bibinfo {author}
  {\bibfnamefont {C.~A.}\ \bibnamefont {Hoffman}}, \bibinfo {author}
  {\bibfnamefont {M.~J.}\ \bibnamefont {Yang}}, \bibinfo {author}
  {\bibfnamefont {B.~R.}\ \bibnamefont {Bennett}}, \bibinfo {author}
  {\bibfnamefont {B.~V.}\ \bibnamefont {Shanabrook}}, \bibinfo {author}
  {\bibfnamefont {H.}~\bibnamefont {Lee}}, \bibinfo {author} {\bibfnamefont
  {C.~H.}\ \bibnamefont {Lin}}, \bibinfo {author} {\bibfnamefont {S.~S.}\
  \bibnamefont {Pei}}, \ and\ \bibinfo {author} {\bibfnamefont {R.~H.}\
  \bibnamefont {Miles}},\ }\href@noop {} {\bibfield  {journal} {\bibinfo
  {journal} {Appl. Phys. Lett.}\ }\textbf {\bibinfo {volume} {73}},\ \bibinfo
  {pages} {2857} (\bibinfo {year} {1998}{\natexlab{a}})}\BibitemShut {NoStop}%
\bibitem [{\citenamefont {Murphy}(2010)}]{murphy2010semiconductor}%
  \BibitemOpen
  \bibfield  {author} {\bibinfo {author} {\bibfnamefont {E.}~\bibnamefont
  {Murphy}},\ }\href@noop {} {\bibfield  {journal} {\bibinfo  {journal} {Nat.
  Photonics}\ }\textbf {\bibinfo {volume} {4}},\ \bibinfo {pages} {287}
  (\bibinfo {year} {2010})}\BibitemShut {NoStop}%
\bibitem [{\citenamefont {Meissner}\ \emph {et~al.}(1991)\citenamefont
  {Meissner}, \citenamefont {Fluegel}, \citenamefont {Binder}, \citenamefont
  {Koch}, \citenamefont {Khitrova},\ and\ \citenamefont
  {Peygambarian}}]{Meissner:1991jm}%
  \BibitemOpen
  \bibfield  {author} {\bibinfo {author} {\bibfnamefont {K.}~\bibnamefont
  {Meissner}}, \bibinfo {author} {\bibfnamefont {B.}~\bibnamefont {Fluegel}},
  \bibinfo {author} {\bibfnamefont {R.}~\bibnamefont {Binder}}, \bibinfo
  {author} {\bibfnamefont {S.~W.}\ \bibnamefont {Koch}}, \bibinfo {author}
  {\bibfnamefont {G.}~\bibnamefont {Khitrova}}, \ and\ \bibinfo {author}
  {\bibfnamefont {N.}~\bibnamefont {Peygambarian}},\ }\href@noop {} {\bibfield
  {journal} {\bibinfo  {journal} {Appl. Phys. Lett.}\ }\textbf {\bibinfo
  {volume} {59}},\ \bibinfo {pages} {259} (\bibinfo {year} {1991})}\BibitemShut
  {NoStop}%
\bibitem [{\citenamefont {Binder}\ \emph {et~al.}(1991)\citenamefont {Binder},
  \citenamefont {Galbraith},\ and\ \citenamefont {Koch}}]{Binder:1991ks}%
  \BibitemOpen
  \bibfield  {author} {\bibinfo {author} {\bibfnamefont {R.}~\bibnamefont
  {Binder}}, \bibinfo {author} {\bibfnamefont {I.}~\bibnamefont {Galbraith}}, \
  and\ \bibinfo {author} {\bibfnamefont {S.~W.}\ \bibnamefont {Koch}},\
  }\href@noop {} {\bibfield  {journal} {\bibinfo  {journal} {Phys. Rev. B}\
  }\textbf {\bibinfo {volume} {44}},\ \bibinfo {pages} {3031} (\bibinfo {year}
  {1991})}\BibitemShut {NoStop}%
\bibitem [{\citenamefont {Galbraith}\ \emph {et~al.}(1992)\citenamefont
  {Galbraith}, \citenamefont {Dawson},\ and\ \citenamefont
  {Foxon}}]{Galbraith:1992gs}%
  \BibitemOpen
  \bibfield  {author} {\bibinfo {author} {\bibfnamefont {I.}~\bibnamefont
  {Galbraith}}, \bibinfo {author} {\bibfnamefont {P.}~\bibnamefont {Dawson}}, \
  and\ \bibinfo {author} {\bibfnamefont {C.~T.}\ \bibnamefont {Foxon}},\
  }\href@noop {} {\bibfield  {journal} {\bibinfo  {journal} {Phys. Rev. B}\
  }\textbf {\bibinfo {volume} {45}},\ \bibinfo {pages} {13499} (\bibinfo {year}
  {1992})}\BibitemShut {NoStop}%
\bibitem [{\citenamefont {Feldmann}\ \emph {et~al.}(1989)\citenamefont
  {Feldmann}, \citenamefont {Sattmann}, \citenamefont {G{\"o}bel},
  \citenamefont {Kuhl}, \citenamefont {Hebling}, \citenamefont {Ploog},
  \citenamefont {Muralidharan}, \citenamefont {Dawson},\ and\ \citenamefont
  {Foxon}}]{Feldmann:1989im}%
  \BibitemOpen
  \bibfield  {author} {\bibinfo {author} {\bibfnamefont {J.}~\bibnamefont
  {Feldmann}}, \bibinfo {author} {\bibfnamefont {R.}~\bibnamefont {Sattmann}},
  \bibinfo {author} {\bibfnamefont {E.~O.}\ \bibnamefont {G{\"o}bel}}, \bibinfo
  {author} {\bibfnamefont {J.}~\bibnamefont {Kuhl}}, \bibinfo {author}
  {\bibfnamefont {J.}~\bibnamefont {Hebling}}, \bibinfo {author} {\bibfnamefont
  {K.}~\bibnamefont {Ploog}}, \bibinfo {author} {\bibfnamefont
  {R.}~\bibnamefont {Muralidharan}}, \bibinfo {author} {\bibfnamefont
  {P.}~\bibnamefont {Dawson}}, \ and\ \bibinfo {author} {\bibfnamefont {C.~T.}\
  \bibnamefont {Foxon}},\ }\href@noop {} {\bibfield  {journal} {\bibinfo
  {journal} {Phys. Rev. Lett.}\ }\textbf {\bibinfo {volume} {62}},\ \bibinfo
  {pages} {1892} (\bibinfo {year} {1989})}\BibitemShut {NoStop}%
\bibitem [{\citenamefont {Berger}\ \emph {et~al.}(2015)\citenamefont {Berger},
  \citenamefont {M{\"o}ller}, \citenamefont {Hens}, \citenamefont {Fuchs},
  \citenamefont {Stolz}, \citenamefont {Koch}, \citenamefont {Ruiz~Perez},
  \citenamefont {Hader},\ and\ \citenamefont {Moloney}}]{Berger:2015ev}%
  \BibitemOpen
  \bibfield  {author} {\bibinfo {author} {\bibfnamefont {C.}~\bibnamefont
  {Berger}}, \bibinfo {author} {\bibfnamefont {C.}~\bibnamefont {M{\"o}ller}},
  \bibinfo {author} {\bibfnamefont {P.}~\bibnamefont {Hens}}, \bibinfo {author}
  {\bibfnamefont {C.}~\bibnamefont {Fuchs}}, \bibinfo {author} {\bibfnamefont
  {W.}~\bibnamefont {Stolz}}, \bibinfo {author} {\bibfnamefont {S.~W.}\
  \bibnamefont {Koch}}, \bibinfo {author} {\bibfnamefont {A.}~\bibnamefont
  {Ruiz~Perez}}, \bibinfo {author} {\bibfnamefont {J.}~\bibnamefont {Hader}}, \
  and\ \bibinfo {author} {\bibfnamefont {J.~V.}\ \bibnamefont {Moloney}},\
  }\href {\doibase 10.1063/1.4917180} {\bibfield  {journal} {\bibinfo
  {journal} {AIP Adv.}\ }\textbf {\bibinfo {volume} {5}},\ \bibinfo {pages}
  {047105} (\bibinfo {year} {2015})}\BibitemShut {NoStop}%
\bibitem [{\citenamefont {Gies}\ \emph {et~al.}(2015)\citenamefont {Gies},
  \citenamefont {Kruska}, \citenamefont {Berger}, \citenamefont {Hens},
  \citenamefont {Fuchs}, \citenamefont {Ruiz~Perez}, \citenamefont {Rosemann},
  \citenamefont {Veletas}, \citenamefont {Chatterjee}, \citenamefont {Stolz},
  \citenamefont {Koch}, \citenamefont {Hader}, \citenamefont {Moloney},\ and\
  \citenamefont {Heimbrodt}}]{Gies:2015ec}%
  \BibitemOpen
  \bibfield  {author} {\bibinfo {author} {\bibfnamefont {S.}~\bibnamefont
  {Gies}}, \bibinfo {author} {\bibfnamefont {C.}~\bibnamefont {Kruska}},
  \bibinfo {author} {\bibfnamefont {C.}~\bibnamefont {Berger}}, \bibinfo
  {author} {\bibfnamefont {P.}~\bibnamefont {Hens}}, \bibinfo {author}
  {\bibfnamefont {C.}~\bibnamefont {Fuchs}}, \bibinfo {author} {\bibfnamefont
  {A.}~\bibnamefont {Ruiz~Perez}}, \bibinfo {author} {\bibfnamefont {N.~W.}\
  \bibnamefont {Rosemann}}, \bibinfo {author} {\bibfnamefont {J.}~\bibnamefont
  {Veletas}}, \bibinfo {author} {\bibfnamefont {S.}~\bibnamefont {Chatterjee}},
  \bibinfo {author} {\bibfnamefont {W.}~\bibnamefont {Stolz}}, \bibinfo
  {author} {\bibfnamefont {S.~W.}\ \bibnamefont {Koch}}, \bibinfo {author}
  {\bibfnamefont {J.}~\bibnamefont {Hader}}, \bibinfo {author} {\bibfnamefont
  {J.~V.}\ \bibnamefont {Moloney}}, \ and\ \bibinfo {author} {\bibfnamefont
  {W.}~\bibnamefont {Heimbrodt}},\ }\href@noop {} {\bibfield  {journal}
  {\bibinfo  {journal} {Appl. Phys. Lett.}\ }\textbf {\bibinfo {volume}
  {107}},\ \bibinfo {pages} {182104} (\bibinfo {year} {2015})}\BibitemShut
  {NoStop}%
\bibitem [{\citenamefont {Klem}\ \emph {et~al.}(2000)\citenamefont {Klem},
  \citenamefont {Blum}, \citenamefont {Kurtz}, \citenamefont {Fritz},\ and\
  \citenamefont {Choquette}}]{Klem}%
  \BibitemOpen
  \bibfield  {author} {\bibinfo {author} {\bibfnamefont {J.~F.}\ \bibnamefont
  {Klem}}, \bibinfo {author} {\bibfnamefont {O.}~\bibnamefont {Blum}}, \bibinfo
  {author} {\bibfnamefont {S.~R.}\ \bibnamefont {Kurtz}}, \bibinfo {author}
  {\bibfnamefont {I.~J.}\ \bibnamefont {Fritz}}, \ and\ \bibinfo {author}
  {\bibfnamefont {K.~D.}\ \bibnamefont {Choquette}},\ }\href@noop {} {\bibfield
   {journal} {\bibinfo  {journal} {J. Vac. Sci. Technol.}\ }\textbf {\bibinfo
  {volume} {18}},\ \bibinfo {pages} {1605} (\bibinfo {year}
  {2000})}\BibitemShut {NoStop}%
\bibitem [{\citenamefont {Meyer}\ \emph
  {et~al.}(1998{\natexlab{b}})\citenamefont {Meyer}, \citenamefont {Olafsen},
  \citenamefont {Aifer}, \citenamefont {Bewley}, \citenamefont {Felix},
  \citenamefont {Vurgaftman}, \citenamefont {Yang}, \citenamefont {Goldberg},
  \citenamefont {Zhang}, \citenamefont {Lin}, \citenamefont {Pei},\ and\
  \citenamefont {Chow}}]{:/content/journals/10.1049/ip-opt_19982304}%
  \BibitemOpen
  \bibfield  {author} {\bibinfo {author} {\bibfnamefont {J.}~\bibnamefont
  {Meyer}}, \bibinfo {author} {\bibfnamefont {L.}~\bibnamefont {Olafsen}},
  \bibinfo {author} {\bibfnamefont {E.}~\bibnamefont {Aifer}}, \bibinfo
  {author} {\bibfnamefont {W.}~\bibnamefont {Bewley}}, \bibinfo {author}
  {\bibfnamefont {C.}~\bibnamefont {Felix}}, \bibinfo {author} {\bibfnamefont
  {V.}~\bibnamefont {Vurgaftman}}, \bibinfo {author} {\bibfnamefont
  {M.}~\bibnamefont {Yang}}, \bibinfo {author} {\bibfnamefont {L.}~\bibnamefont
  {Goldberg}}, \bibinfo {author} {\bibfnamefont {D.}~\bibnamefont {Zhang}},
  \bibinfo {author} {\bibfnamefont {C.-H.}\ \bibnamefont {Lin}}, \bibinfo
  {author} {\bibfnamefont {S.}~\bibnamefont {Pei}}, \ and\ \bibinfo {author}
  {\bibfnamefont {D.}~\bibnamefont {Chow}},\ }\href
  {http://digital-library.theiet.org/content/journals/10.1049/ip-opt_19982304}
  {\bibfield  {journal} {\bibinfo  {journal} {IEE Proc. - Optoelectron.}\
  }\textbf {\bibinfo {volume} {145}},\ \bibinfo {pages} {275} (\bibinfo {year}
  {1998}{\natexlab{b}})}\BibitemShut {NoStop}%
\bibitem [{\citenamefont {Hakki}(1973)}]{Hakki:1973fm}%
  \BibitemOpen
  \bibfield  {author} {\bibinfo {author} {\bibfnamefont {B.~W.}\ \bibnamefont
  {Hakki}},\ }\href@noop {} {\bibfield  {journal} {\bibinfo  {journal} {J. of
  Appl. Phys.}\ }\textbf {\bibinfo {volume} {44}},\ \bibinfo {pages} {4113}
  (\bibinfo {year} {1973})}\BibitemShut {NoStop}%
\bibitem [{\citenamefont {Shterengas}\ \emph {et~al.}(2013)\citenamefont
  {Shterengas}, \citenamefont {Liang}, \citenamefont {Kipshidze}, \citenamefont
  {Hosoda}, \citenamefont {Suchalkin},\ and\ \citenamefont
  {Belenky}}]{Shterengas:2013dl}%
  \BibitemOpen
  \bibfield  {author} {\bibinfo {author} {\bibfnamefont {L.}~\bibnamefont
  {Shterengas}}, \bibinfo {author} {\bibfnamefont {R.}~\bibnamefont {Liang}},
  \bibinfo {author} {\bibfnamefont {G.}~\bibnamefont {Kipshidze}}, \bibinfo
  {author} {\bibfnamefont {T.}~\bibnamefont {Hosoda}}, \bibinfo {author}
  {\bibfnamefont {S.}~\bibnamefont {Suchalkin}}, \ and\ \bibinfo {author}
  {\bibfnamefont {G.}~\bibnamefont {Belenky}},\ }\href@noop {} {\bibfield
  {journal} {\bibinfo  {journal} {Appl. Phys. Lett.}\ }\textbf {\bibinfo
  {volume} {103}},\ \bibinfo {pages} {121108} (\bibinfo {year}
  {2013})}\BibitemShut {NoStop}%
\bibitem [{\citenamefont {Shaklee}\ \emph {et~al.}(1973)\citenamefont
  {Shaklee}, \citenamefont {Nahory},\ and\ \citenamefont
  {Leheny}}]{Shaklee:1973hi}%
  \BibitemOpen
  \bibfield  {author} {\bibinfo {author} {\bibfnamefont {K.~L.}\ \bibnamefont
  {Shaklee}}, \bibinfo {author} {\bibfnamefont {R.~E.}\ \bibnamefont {Nahory}},
  \ and\ \bibinfo {author} {\bibfnamefont {R.~F.}\ \bibnamefont {Leheny}},\
  }\href@noop {} {\bibfield  {journal} {\bibinfo  {journal} {JOL}\ }\textbf
  {\bibinfo {volume} {7}},\ \bibinfo {pages} {284} (\bibinfo {year}
  {1973})}\BibitemShut {NoStop}%
\bibitem [{\citenamefont {Kirstaedter}\ \emph {et~al.}(1996)\citenamefont
  {Kirstaedter}, \citenamefont {Schmidt}, \citenamefont {Ledentsov},
  \citenamefont {Bimberg}, \citenamefont {Ustinov}, \citenamefont {Egorov},
  \citenamefont {Zhukov}, \citenamefont {Maximov}, \citenamefont
  {Kop{\textquoteright}ev},\ and\ \citenamefont
  {Alferov}}]{Kirstaedter:1996fe}%
  \BibitemOpen
  \bibfield  {author} {\bibinfo {author} {\bibfnamefont {N.}~\bibnamefont
  {Kirstaedter}}, \bibinfo {author} {\bibfnamefont {O.~G.}\ \bibnamefont
  {Schmidt}}, \bibinfo {author} {\bibfnamefont {N.~N.}\ \bibnamefont
  {Ledentsov}}, \bibinfo {author} {\bibfnamefont {D.}~\bibnamefont {Bimberg}},
  \bibinfo {author} {\bibfnamefont {V.~M.}\ \bibnamefont {Ustinov}}, \bibinfo
  {author} {\bibfnamefont {A.~Y.}\ \bibnamefont {Egorov}}, \bibinfo {author}
  {\bibfnamefont {A.~E.}\ \bibnamefont {Zhukov}}, \bibinfo {author}
  {\bibfnamefont {M.~V.}\ \bibnamefont {Maximov}}, \bibinfo {author}
  {\bibfnamefont {P.~S.}\ \bibnamefont {Kop{\textquoteright}ev}}, \ and\
  \bibinfo {author} {\bibfnamefont {Z.~I.}\ \bibnamefont {Alferov}},\
  }\href@noop {} {\bibfield  {journal} {\bibinfo  {journal} {Appl. Phys.
  Lett.}\ }\textbf {\bibinfo {volume} {69}},\ \bibinfo {pages} {1226} (\bibinfo
  {year} {1996})}\BibitemShut {NoStop}%
\bibitem [{\citenamefont {Hall}\ \emph {et~al.}(1990)\citenamefont {Hall},
  \citenamefont {Mark}, \citenamefont {Ippen},\ and\ \citenamefont
  {Eisenstein}}]{Hall:1990dv}%
  \BibitemOpen
  \bibfield  {author} {\bibinfo {author} {\bibfnamefont {K.~L.}\ \bibnamefont
  {Hall}}, \bibinfo {author} {\bibfnamefont {J.}~\bibnamefont {Mark}}, \bibinfo
  {author} {\bibfnamefont {E.~P.}\ \bibnamefont {Ippen}}, \ and\ \bibinfo
  {author} {\bibfnamefont {G.}~\bibnamefont {Eisenstein}},\ }\href@noop {}
  {\bibfield  {journal} {\bibinfo  {journal} {Appl. Phys. Lett.}\ }\textbf
  {\bibinfo {volume} {56}},\ \bibinfo {pages} {1740} (\bibinfo {year}
  {1990})}\BibitemShut {NoStop}%
\bibitem [{\citenamefont {Giessen}\ \emph {et~al.}(1996)\citenamefont
  {Giessen}, \citenamefont {Woggon}, \citenamefont {Fluegel}, \citenamefont
  {Mohs}, \citenamefont {Hu}, \citenamefont {Koch},\ and\ \citenamefont
  {Peyghambarian}}]{Giessen:1996db}%
  \BibitemOpen
  \bibfield  {author} {\bibinfo {author} {\bibfnamefont {H.}~\bibnamefont
  {Giessen}}, \bibinfo {author} {\bibfnamefont {U.}~\bibnamefont {Woggon}},
  \bibinfo {author} {\bibfnamefont {B.}~\bibnamefont {Fluegel}}, \bibinfo
  {author} {\bibfnamefont {G.}~\bibnamefont {Mohs}}, \bibinfo {author}
  {\bibfnamefont {Y.~Z.}\ \bibnamefont {Hu}}, \bibinfo {author} {\bibfnamefont
  {S.~W.}\ \bibnamefont {Koch}}, \ and\ \bibinfo {author} {\bibfnamefont
  {N.}~\bibnamefont {Peyghambarian}},\ }\href@noop {} {\bibfield  {journal}
  {\bibinfo  {journal} {Opt. Lett.}\ }\textbf {\bibinfo {volume} {21}},\
  \bibinfo {pages} {1043} (\bibinfo {year} {1996})}\BibitemShut {NoStop}%
\bibitem [{\citenamefont {Lange}\ \emph {et~al.}(2007)\citenamefont {Lange},
  \citenamefont {Chatterjee}, \citenamefont {Schlichenmaier}, \citenamefont
  {Thr{\"a}nhardt}, \citenamefont {Koch}, \citenamefont {R{\"u}hle},
  \citenamefont {Hader}, \citenamefont {Moloney}, \citenamefont {Khitrova},\
  and\ \citenamefont {Gibbs}}]{Lange:2007ek}%
  \BibitemOpen
  \bibfield  {author} {\bibinfo {author} {\bibfnamefont {C.}~\bibnamefont
  {Lange}}, \bibinfo {author} {\bibfnamefont {S.}~\bibnamefont {Chatterjee}},
  \bibinfo {author} {\bibfnamefont {C.}~\bibnamefont {Schlichenmaier}},
  \bibinfo {author} {\bibfnamefont {A.}~\bibnamefont {Thr{\"a}nhardt}},
  \bibinfo {author} {\bibfnamefont {S.~W.}\ \bibnamefont {Koch}}, \bibinfo
  {author} {\bibfnamefont {W.~W.}\ \bibnamefont {R{\"u}hle}}, \bibinfo {author}
  {\bibfnamefont {J.}~\bibnamefont {Hader}}, \bibinfo {author} {\bibfnamefont
  {J.~V.}\ \bibnamefont {Moloney}}, \bibinfo {author} {\bibfnamefont
  {G.}~\bibnamefont {Khitrova}}, \ and\ \bibinfo {author} {\bibfnamefont
  {H.~M.}\ \bibnamefont {Gibbs}},\ }\href@noop {} {\bibfield  {journal}
  {\bibinfo  {journal} {Appl. Phys. Lett.}\ }\textbf {\bibinfo {volume} {90}},\
  \bibinfo {pages} {251102} (\bibinfo {year} {2007})}\BibitemShut {NoStop}%
\bibitem [{\citenamefont {Lindberg}\ and\ \citenamefont
  {Koch}(1988)}]{Lindberg:1988hf}%
  \BibitemOpen
  \bibfield  {author} {\bibinfo {author} {\bibfnamefont {M.}~\bibnamefont
  {Lindberg}}\ and\ \bibinfo {author} {\bibfnamefont {S.~W.}\ \bibnamefont
  {Koch}},\ }\href {\doibase 10.1103/PhysRevB.38.3342} {\bibfield  {journal}
  {\bibinfo  {journal} {Phys. Rev. B}\ }\textbf {\bibinfo {volume} {38}},\
  \bibinfo {pages} {3342} (\bibinfo {year} {1988})}\BibitemShut {NoStop}%
\bibitem [{\citenamefont {Haug}\ and\ \citenamefont
  {Koch}(2009)}]{Haug:2009tb}%
  \BibitemOpen
  \bibfield  {author} {\bibinfo {author} {\bibfnamefont {H.}~\bibnamefont
  {Haug}}\ and\ \bibinfo {author} {\bibfnamefont {S.~W.}\ \bibnamefont
  {Koch}},\ }\href@noop {} {\emph {\bibinfo {title} {{Quantum Theory of the
  Optical and Electronic Properties of Semiconductors}}}},\ \bibinfo {edition}
  {5th}\ ed.\ (\bibinfo  {publisher} {World Scientific Publ.},\ \bibinfo
  {address} {Singapore},\ \bibinfo {year} {2009})\BibitemShut {NoStop}%
\bibitem [{\citenamefont {Kira}\ and\ \citenamefont
  {Koch}(2012)}]{Kira:2011up}%
  \BibitemOpen
  \bibfield  {author} {\bibinfo {author} {\bibfnamefont {M.}~\bibnamefont
  {Kira}}\ and\ \bibinfo {author} {\bibfnamefont {S.~W.}\ \bibnamefont
  {Koch}},\ }\href@noop {} {\emph {\bibinfo {title} {{Semiconductor Quantum
  Optics}}}}\ (\bibinfo  {publisher} {Cambridge University Press},\ \bibinfo
  {address} {Cambridge},\ \bibinfo {year} {2012})\BibitemShut {NoStop}%
\bibitem [{\citenamefont {Chow}\ and\ \citenamefont
  {Koch}(1999)}]{Chow:1999ts}%
  \BibitemOpen
  \bibfield  {author} {\bibinfo {author} {\bibfnamefont {W.~W.}\ \bibnamefont
  {Chow}}\ and\ \bibinfo {author} {\bibfnamefont {S.~W.}\ \bibnamefont
  {Koch}},\ }\href@noop {} {\emph {\bibinfo {title} {{Semiconductor-Laser
  Fundamentals: Physics of the Gain Materials}}}}\ (\bibinfo  {publisher}
  {Springer},\ \bibinfo {address} {Berlin, Heidelberg, New York},\ \bibinfo
  {year} {1999})\BibitemShut {NoStop}%
\bibitem [{\citenamefont {Hader}\ \emph
  {et~al.}(2003{\natexlab{a}})\citenamefont {Hader}, \citenamefont {Koch},\
  and\ \citenamefont {Moloney}}]{Hader:2003hi}%
  \BibitemOpen
  \bibfield  {author} {\bibinfo {author} {\bibfnamefont {J.}~\bibnamefont
  {Hader}}, \bibinfo {author} {\bibfnamefont {S.~W.}\ \bibnamefont {Koch}}, \
  and\ \bibinfo {author} {\bibfnamefont {J.~V.}\ \bibnamefont {Moloney}},\
  }\href {\doibase 10.1016/S0038-1101(02)00405-7} {\bibfield  {journal}
  {\bibinfo  {journal} {Solid-State Electron.}\ }\textbf {\bibinfo {volume}
  {47}},\ \bibinfo {pages} {513} (\bibinfo {year}
  {2003}{\natexlab{a}})}\BibitemShut {NoStop}%
\bibitem [{\citenamefont {Hader}\ \emph
  {et~al.}(2003{\natexlab{b}})\citenamefont {Hader}, \citenamefont {Moloney},
  \citenamefont {Koch},\ and\ \citenamefont {Chow}}]{Hader:2003hm}%
  \BibitemOpen
  \bibfield  {author} {\bibinfo {author} {\bibfnamefont {J.}~\bibnamefont
  {Hader}}, \bibinfo {author} {\bibfnamefont {J.~V.}\ \bibnamefont {Moloney}},
  \bibinfo {author} {\bibfnamefont {S.~W.}\ \bibnamefont {Koch}}, \ and\
  \bibinfo {author} {\bibfnamefont {W.~W.}\ \bibnamefont {Chow}},\ }\href
  {\doibase 10.1109/JSTQE.2003.818342} {\bibfield  {journal} {\bibinfo
  {journal} {{IEEE} J. Select. Topics Quantum Electron.}\ }\textbf {\bibinfo
  {volume} {9}},\ \bibinfo {pages} {688} (\bibinfo {year}
  {2003}{\natexlab{b}})}\BibitemShut {NoStop}%
\bibitem [{\citenamefont {Hader}\ \emph {et~al.}(1997)\citenamefont {Hader},
  \citenamefont {Linder},\ and\ \citenamefont {D{\"o}hler}}]{Hader:1997gr}%
  \BibitemOpen
  \bibfield  {author} {\bibinfo {author} {\bibfnamefont {J.}~\bibnamefont
  {Hader}}, \bibinfo {author} {\bibfnamefont {N.}~\bibnamefont {Linder}}, \
  and\ \bibinfo {author} {\bibfnamefont {G.~H.}\ \bibnamefont {D{\"o}hler}},\
  }\href {\doibase 10.1103/PhysRevB.55.6960} {\bibfield  {journal} {\bibinfo
  {journal} {Phys. Rev. B}\ }\textbf {\bibinfo {volume} {55}},\ \bibinfo
  {pages} {6960} (\bibinfo {year} {1997})}\BibitemShut {NoStop}%
\bibitem [{\citenamefont {Ahn}\ and\ \citenamefont
  {Chuang}(1988)}]{Ahn:1988ks}%
  \BibitemOpen
  \bibfield  {author} {\bibinfo {author} {\bibfnamefont {D.}~\bibnamefont
  {Ahn}}\ and\ \bibinfo {author} {\bibfnamefont {S.~L.}\ \bibnamefont
  {Chuang}},\ }\href {\doibase 10.1063/1.342118} {\bibfield  {journal}
  {\bibinfo  {journal} {J. Appl. Phys.}\ }\textbf {\bibinfo {volume} {64}},\
  \bibinfo {pages} {6143} (\bibinfo {year} {1988})}\BibitemShut {NoStop}%
\bibitem [{\citenamefont {Born}\ and\ \citenamefont
  {Wolf}(1999)}]{Born:1999un}%
  \BibitemOpen
  \bibfield  {author} {\bibinfo {author} {\bibfnamefont {M.}~\bibnamefont
  {Born}}\ and\ \bibinfo {author} {\bibfnamefont {E.}~\bibnamefont {Wolf}},\
  }\href@noop {} {\emph {\bibinfo {title} {{Principles of Optics}}}},\ \bibinfo
  {edition} {7th}\ ed.,\ Electromagnetic Theory of Propagation, Interference
  and Diffraction of Light\ (\bibinfo  {publisher} {Cambridge University
  Press},\ \bibinfo {address} {Cambridge},\ \bibinfo {year} {1999})\BibitemShut
  {NoStop}%
\bibitem [{\citenamefont {Schafer}\ \emph {et~al.}(2008)\citenamefont
  {Schafer}, \citenamefont {Hoyer}, \citenamefont {Kira}, \citenamefont
  {Koch},\ and\ \citenamefont {Moloney}}]{Schafer:2008in}%
  \BibitemOpen
  \bibfield  {author} {\bibinfo {author} {\bibfnamefont {M.}~\bibnamefont
  {Schafer}}, \bibinfo {author} {\bibfnamefont {W.}~\bibnamefont {Hoyer}},
  \bibinfo {author} {\bibfnamefont {M.}~\bibnamefont {Kira}}, \bibinfo {author}
  {\bibfnamefont {S.~W.}\ \bibnamefont {Koch}}, \ and\ \bibinfo {author}
  {\bibfnamefont {J.~V.}\ \bibnamefont {Moloney}},\ }\href {\doibase
  10.1364/JOSAB.25.000187} {\bibfield  {journal} {\bibinfo  {journal} {J. Opt.
  Soc. Am. B}\ }\textbf {\bibinfo {volume} {25}},\ \bibinfo {pages} {187}
  (\bibinfo {year} {2008})}\BibitemShut {NoStop}%
\end{thebibliography}%

\end{document}